\documentclass[runningheads,a4paper]{llncs}

\usepackage{graphicx}

\usepackage{url}
\usepackage{cite}

\usepackage{booktabs} 
\usepackage{multirow, array}
\usepackage{pbox}
\usepackage{algorithmic}
\usepackage{array}
\usepackage{fixltx2e}
\usepackage{stfloats}
\usepackage{url}
\usepackage{algorithm}
\usepackage{hyperref}
\usepackage[english]{babel} 
\usepackage[sharp]{easylist}
\usepackage[english]{babel}
\usepackage[utf8]{inputenc} 
\usepackage{subfig}
\usepackage{textcomp}
\usepackage{threeparttable}
\usepackage[dotinlabels]{titletoc}
\usepackage{tabulary}
\usepackage{tabularx}
\usepackage{lipsum}
\usepackage{xfrac}
\usepackage{amsmath}
\usepackage{paralist}
\usepackage{verbatim}
\usepackage{enumerate}
\usepackage{amstext}

\urldef{\mailsa}\path|{lorena.recalde,ricardo.baeza}@upf.edu|
\newcommand{\keywords}[1]{\par\addvspace\baselineskip
\noindent\keywordname\enspace\ignorespaces#1}

\begin{document}

\mainmatter  

\title{What kind of content are you prone to tweet?\\
Multi-topic Preference Model for Tweeters}
\titlerunning{Multi-topic Preference Model for Tweeters}

%
%
%

\author{Lorena Recalde \and Ricardo Baeza-Yates}
\authorrunning{Multi-topic Preference Model for Tweeters}

\institute{Web Science and Social Computing Research Group\\Department of Information and Communication Technologies\\Universitat Pompeu Fabra\\
Barcelona, 08018 Spain\\
\mailsa}

%
%

\maketitle

\begin{abstract}

According to tastes, a person could show preference for a given category of content to a greater or lesser extent. However, quantifying people's amount of interest in a certain topic is a challenging task, especially considering the massive digital information they are exposed to. For example, in the context of Twitter, aligned with his/her preferences a user may tweet and retweet more about technology than sports and do not share any music-related content. The problem we address in this paper is the identification of users' implicit topic preferences by analyzing the content categories they tend to post on Twitter. Our proposal is significant given that modeling their multi-topic profile may be useful to find patterns or association between preferences for categories, discover trending topics and cluster similar users to generate better group recommendations of content. In the present work, we propose a method based on the Mixed Gaussian Model to extract the multidimensional preference representation for 399 Ecuadorian tweeters concerning twenty-two different topics (or dimensions) which became known by manually categorizing 68.186 tweets. Our experiment findings indicate that the proposed approach is effective at detecting the topic interests of users.

\keywords{Multidimensional Profile; User Modeling; Expectation Maximization; Group Recommender System, Topic Modeling, Twitter.}
\end{abstract}

\section{Introduction}
In the light of the massive digital information people are exposed to, they show interest in diverse topics to a greater or lesser extent. Quantifying and measuring a user's degree of interest in certain content and finding its correlation with his/her preference for another topic is a challenging task, especially in social media platforms where the user interests are not static. For example, people highly engaged to culture-related topics may often retweet posts about next concerts, but when their favorite soccer team wins a match, they generate posts according to that. Therefore, identifying this kind of topic preferences association represented as a multidimensional user model, (\textit{MUM}), may be meaningful to define how much the user shows interest in content categories as well as to group like-minded users and address better recommendations for them. 

In the context of Twitter, automatically classifying a tweet into a topic category is hard to achieve. Indeed, having a group of words that form a sentence of less than 140 characters\footnote{When the dataset was collected Twitter posts were limited to 140 characters. Currently, the length of a tweet may be up 280 characters.} and that contains abbreviations, emoticons, URLs and mentions of other users, which in particular do not provide a relevant meaning by themselves, makes the semantic analysis a challenge. Then, during the classification work of a tweet, the capture of other words like hashtags, proper nouns, compound nouns and verbs lead to a better topic assignment. Accordingly, to make the implementation of the comprehension and classification tasks of a tweet possible (as the basic step to then associate topic interest to tweeters) we propose a method that merges language modeling techniques and the Expectation Maximization algorithm \cite{10.2307/2984875} (\textit{EM} for Mixture of Gaussians). The strategy is independent from the users' posts language which makes it feasible to take Spanish tweets posted by Ecuadorians as our case study. Respectively, aggregating the Mixed Gaussian Model (topic soft assignments) of the target users' tweets in order to find their \textit{MUM}  is useful to cluster them and find groups of users interested in the same topics and to the same extent. 

There are loads of research works in the field of users' topic preferences modeling. However, to the best of our knowledge, our proposal represents the first attempt to quantify the degree of responsibility a topic has over a given tweeter. That is to say, the method allows to identify the percentage in which each category takes part in the user profile. Given this real-world application scenario, our scientific contributions are:

\begin{compactitem}
\item a method to define the multidimensional user model. \textit{MUM} for tweeters, which can be further applied to cluster like-minded users and design group recommendations;
\item an evaluation of the accuracy of the proposed method considering, in terms of a comparative analysis, a baseline approach which takes a \textit{ground-truth dataset} of labeled tweets. In such way, the \textit{MUM} approach is compared to the results of a traditional machine learning classifier;
\item a detailed validation of our approach that shows its effectiveness in modeling users. We show that similar tweeters, whose profiles were modeled with MUM, are able to be grouped together.
\end{compactitem}

In summary, in this paper we propose a novel method for unsupervised and topic-based ``soft'' classification of tweets. Such approach is used to model Twitter users. The remainder of the paper is organized as follows. Section~\ref{related} summarizes the context of the present research and related literature; moreover, we draw a comparison to our proposal; Section~\ref{approach} describes our approach; in Section~\ref{experiments} we present the experimental framework and the obtained results. Finally, some conclusions and issues about future work are explained in Section~\ref{conclusions}.

%

\section{Context and Related Work}\label{related}
Human factors such as need for approval, acceptance of a community, reputation as an expert, friendship, among others are implicitly present in Online Social Networks, OSNs \cite{Grabner2009}. 
Few of these factors have settled in a specific social media with more intensity than others, and human curiosity satisfaction is a widespread one. For example, \textit{curiosity} to know about acquaintances' activities is prevalent in Facebook; on the other hand, \textit{curiosity} to know (and learn) about new content related to one's topics of interest is seen in Twitter. Therefore, to meet user's curiosity it is necessary to present them with others' posts that are certainly of their preference. Modeling users' profiles is essential to find the topics they enjoy consuming and provide the curious users with meaningful information. Accordingly, in this section we present related works considering \textit{Tweeters Modeling for Recommender Systems} whose aim is to link tweeters with the corresponding content/items. Later, \textit{Group Formation and Group Recommendation} is detailed due to the further application of our approach in this area. Finally, as our proposal is based on the use of EM to find the \textit{degree of responsibility} a topic has over a tweet, \textit{Tweets Classification} works are also described.
\subsection{Tweeters Modeling for Recommendation}
Recommender systems predict if an \textit{unseen item} is going to be of interest of a target user. To address the problem of recommendation in the Social Web such systems mine people's interactions, trust connections, previously adopted suggestions, use of self-annotated content (\textit{i.e.} through hashtags), groups subscription, among others.
Tweet recommendation has been studied due to the constant threat of content overload in the users time-line. In \cite{Chen:2012:CPT:2348283.2348372}, the approach makes use of three components: tweet topic level factors, user social relation factors and explicit features like authority of the tweet creator and quality of the tweet to define if a tweet can be recommended. Unlike our proposal, this article bases the user model in the social connections and not in topics of interest. Research presented in \cite{Chen:2010:STE:1753326.1753503} proposes a URLs recommender system for tweeters based on content sources, topic interest models of users, and social voting. Their findings show that topic relevance and social interactions were helpful in presenting recommendations. As in our approach, \cite{Chen:2010:STE:1753326.1753503} builds the user's profile from his/her own tweets. However, they work with the weighting scheme {tf-idf} \cite{Salton:1988:TAA:54259.54260} to find the relevant topics for the user while we apply word embeddings.   
\subsection{Groups Formation and Recommendations}
From a general perspective, the benefits of using a microblogging platform such as Twitter emerge from the activity of the users themselves. This social and data-oriented phenomenon is known as collective intelligence \cite{Lai2008, Surowiecki:2005}. 
For example, a recommender system that tracks events liked by the users may infer that the users who attend musicals twice a month also attend plays once a month. This generalization may be done because the system learns patterns from the behavior of the whole community. In such a case, like-minded users need to be grouped and analyzed together. A Group Recommender System supports the recommendation process by using aggregation methods in order to model the preferences of a group of people. This is needed when there is an activity (domain) that can be done or enjoyed in groups \cite{boratto1}. For our proposal, it may be possible to detect groups of tweeters interested in the same topics and suggest for them, for example, lists to subscribe in.  

\subsection{Tweets Classification}
In terms of tweets classification, in \cite{ Sriram:2010:STC:1835449.1835643}, 5 content categories (News, Events, Opinions, Deals, and Private Messages) are proposed in order to classify short text. In this work, tweets are modeled considering 8 specific features which lead to determine the class of a tweet. For example, one of the features is \textit{presence of time-event phrases} that, in case it is true for a given tweet, might relate it to the Events category. On the other hand, considering the feature \textit{presence of slang words, shortenings} as true for the tweet suggests a Private Message class. While, this method works with more general categories and a supervised classifier, our proposal allows a 300-dimension representation of tweets which are later classified (with soft assignments) considering 22 categories.

In \cite{Godin:2013:UTM:2487788.2488002}, the problem of hashtag prediction is investigated to recommend the users proper hashtags for their tweets. As a first step, Naïve Bayes and the Expectation Maximization algorithm are employed to classify English and non-English tweets. Later, LDA with Gibbs sampling is applied to find the tweet-topic distribution. Like our proposal, EM was employed as a means of unsupervised classification of tweets. However, we used it to model the tweets depending on the hidden topics, to then seeing the tweet model as a percentage allocation per topic. On the other hand, the mentioned work uses EM to identify the probability of a tweet as being writing in English (it results in a hard class assignment).

Topic modeling has been broadly used as means of tweet classification. In \cite{Iman2017ALS}, the authors propose a method where a group of four classifiers are trained to learn the topics for tweet categorization. They define ten topics and with the help of annotators, they classify a set of hashtags into those topics. Once the hashtags are classified, they can label tweets (containing the hashtags) with the corresponding topic. In their experiments they try to find the features and feature classes relevant to maximize the topic classification performance. The baseline method employed to validate our approach follows the same strategy in terms of supervised classification. In \cite{Yang:2014}, a real-time high-precision tweet topic modeling system is proposed. 300 topics are considered, and the proposal is based on an integrative inference algorithm trough supervised learning as well. In contrast, we present a method to categorize tweets in an unsupervised manner. Our method is effective in calculating the degree of participation of a topic in a given tweet (soft clustering) and no labeled data is required.

\section{Approach}\label{approach}
In this section we present the core phases that were implemented to \textit{i)} identify the level of participation or responsibility that each category has over a tweet and \textit{ii)} aggregate the user's tweets classification extracted in the former phase to then define his/her multidimensional user model \textit{MUM}. The \textit{MUM} approach, consists of:

\begin{enumerate}
    \item {\bf Tweets Modeling.} By using word2vec {\cite{mikolov2013distributed}}, we find a vector representation for a given tweet.
    \item {\bf Extraction of the suitable Number of Topics.} A widely known technique to define the number of topics hidden in a corpus is the Elbow method {\cite{Thorndike53whobelongs}}. We use it to decide how many dimensions our tweet/user model will have.
    \item {\bf Tweets Classification.} To define the topics' responsibility degree over a tweet we use EM. As a result, every tweet will have a vector with $K$ dimensions where $K$ depends on the number of topics. Every feature value of the vector is the percentage of the participation of the corresponding topic in the given tweet.
    \item {\bf Twitter Users Model.} Once the strategy to model a tweet is established as formulated in the previous phase, it is applied to the tweets of the target user. We aggregate the results to define the multidimensional user model.
    \item {\bf Grouping like-minded Users.} $MUM$ provides a profile of tweeters who may be clustered in groups of homogeneous interests.
\end{enumerate}

What follows presents the details of our approach considering each task.

\subsection{Tweets Modeling}\label{modeling}
A collection of tweets is employed to build a vector representation model for the words (vocabulary). We use a word embedding strategy based on a neural language model, {\em word2vec}, and its implementation {\em skip-gram}. The model learns to map each word into a low-dimensional continuous vector-space from its distributional properties observed in the provided corpus of tweets\footnote{https://code.google.com/archive/p/word2vec/}. To train the model, a file that contains a tweet per row is needed. 

Other input parameters have to be provided: {\em size} or number of vector dimensions, {\em window} or maximum skip length between words, {\em sample} or threshold for how often the words occur, and {\em min\_count} or minimum number of times a word must occur to be considered. The output of the trained model is a vector for each word in the corpus. Since the vectors are linear, we can sum several vectors to obtain a unique model representation (additive compositionality property). Therefore, in order to create a model of a \textit{tweet} from the words in it, we sum its words vectors. Let $W_t$ be the set of words in the considered tweet $t$. By taking their embeddings, $w_t$ being the vector for a given word, we build the tweet model as follows:
\begin{equation}\label{tweet_vector}
w'_t=\sum_{w_t\in W_t}{w_t}
\end{equation}

Then, the vector representation for $t$ is $w'_t$.
\\
The authors of this paper have worked in tweets modeling with word2vec in previous research projects, and the detailed methodology which covers tweets cleaning/pre-processing and text modeling is explained in \cite{Recalde:2017}. It is worth mentioning that the tweets are being represented as 300-dimension vectors. The values that the parameters took in this study are reported in the section \ref{strategy} to ensure the repeatability of the experiments.

\subsection{Extraction of the suitable Number of Topics}\label{suitable}
To define the number of topics in which tweeters tend to get involved, we take the $w'_t$ or tweets representation extracted previously and try to find the appropriate number of clusters of tweets. Therefore, we may find a meaningful topic per cluster by inspecting the tweets in it (in case the clusters need to be labeled). To separate the tweets into clusters, we applied \textit{K-Means++} \cite{Arthur:2007}. This method spreads out the initial set of cluster centroids, so that they are not too close together. By applying \textit{K-Means++}, it is possible to find an optimal set of centroids, which is required to have optimal means to initialize {EM}.  

The intuition behind clustering is that objects within a cluster are as similar as possible, whereas objects from different clusters are as dissimilar as possible. However, the optimal clustering is somehow subjective and dependent of the final purpose of the clusters; that is to say, the level of detail required from the partitions. The clusters we obtain may suffer from a wide variation of the number of samples in each cluster (\textit{e.g.} few tweets talking about religion and lots talking about politics) so the distribution is not normal. Nevertheless, we can select the number of clusters by using the heterogeneity convergence metric as the \textit{Elbow} method specifies. We are required to run tests considering different $K$ values (\textit{i.e. number of clusters}). To measure distances we use the cosine distance metric. Then, having $K$, we measure the intra-cluster distances between $n$ points in a given cluster $C_k$ and the centroid $c_C$ of that cluster. 

\begin{equation*}
D_k=\sum_{i=1}^n cosineDistance({x}_i, c_C)^2  \quad {x}_i \in C_k \quad \wedge \quad n=|C_k|
\end{equation*}

Finally, adding the intra-cluster sums of squares gives a measure of the compactness of the clustering:
\begin{equation}\label{heterogeneity}
het_k = \sum_{k=1}^K D_k
\end{equation}
In the \textit{Elbow} heuristic we need to visualize the curve by plotting the heterogeneity value $het_k$ against the number of clusters $K$. At certain point, the gain will drop, forming an angle in the graph. Therefore, the graph where we have the heterogeneity versus $K$ allows us to look for the ``Elbow" of the curve where the heterogeneity decreases \textit{rapidly} before this value of $K$, but then only \textit{gradually} for larger values of $K$. The details of the analysis for the case of our study are presented in the experimental setup (Section \ref{strategy}).

While doing the experiments with different $K$ values, we need to keep track not only the heterogeneity (used to apply the Elbow method), but also the centroids $c_C$ calculated for the clusters. 

\subsection{Tweets Classification: the EM algorithm applied over tweets}\label{em}
When the number of topics, specified by the number of clusters found in the previous phase is obtained, the next step is the implementation of the Expectation Maximization algorithm. EM is sensitive to the choice of initial means. With a bad initial set of means, EM might generate clusters that span a large area and are mostly overlapping. Then, instead of initializing means by selecting random points, we take the final set of centroids calculated before (suitable set of initial means). Indeed, the initialization values for EM will be: \textit{i)} initial means, the cluster centroids $c_C$ extracted for the chosen $K$; \textit{ii)} initial weights, we will initialize each cluster weight as the proportion of tweets assigned by k-means++ to that cluster $C_k$; in other words, $n/N$ for $n=|C_k|$ and $N=$ total number of tweets; iii) initial covariance matrix, to initialize the covariance parameters, we compute $\sum_{i=1}^N (x_{ij} - \mu_{C_kj})^2$ for each dimension $j$.

When the initial parameters are set, the input for the algorithm will be the vectors which belong to the tweets that we want to model. The EM algorithm will be in charge of defining the degree of responsibility the topics will have over each tweet. Then, the output after running the algorithm will be the \textit{responsibility matrix}\footnote{Refer to the repository \url{https://github.com/lore10/Multidimensional_User_Profile} to access the code related to the EM algorithm (datasets and other files are also included).} which cardinality is $N$x$K$. The rows of the matrix specify in which extent the observation $x_{i}$ was assigned to the different $K$ topics (columns). In other words, if the topic 0 (or cluster 0) has full responsibility over the observation the value is going to be 1. If we see shared responsibility between eight topics over another tweet, the sum of those values will be 1 (Refer to Section \ref{strategy} to see an example).

\subsection{Twitter Users Model: extraction of the $MUM$}\label{mum}
Having the responsibility matrix, we need to identify which tweets (rows of the matrix) correspond to the given user (noting $t$ as a modeled tweet $\in T_u$). Whence, for the user being analyzed we will have a $|T_u|$x$K$ submatrix, which will be noted as $U$. To establish the user model, we apply next equations.
\begin{equation}\label{userEq1}
sum_j = \sum_{i=0}^{|T_u|-1} t_{ij} 
\end{equation}

For $j \in [0, K-1]$.
Then, we sum the vector values $j$ to obtain the total:
\begin{equation}\label{userEq2}
total = \sum_{j=0}^{K-1} sum_j
\end{equation}
Finally, the model for the user will be represented as percentages:
\begin{equation}\label{userEq3}
MUM_j = (sum_j/total)*100
\end{equation}
In conclusion, {MUM} is going to be a vector of $K$ dimensions that models the given user according to the topics he/she tends to tweet about. The $j$  values will express the extent of topic participation in the user's Twitter profile.

\subsection{Grouping like-minded Users}\label{clustering}
One of the applications of the multi-topic model of users would be clustering similar users to analyze audiences on Twitter, targeting certain groups of tweeters with recommendations, studying subtopics of interest given a group, among others. In the case of our study, this step was taken to evaluate the proposed approach performance. The clustering algorithm we used was K-means++ \cite{Macqueen67somemethods}, which implementation is provided in the tool Graphlab \cite{graphlab} for Python (K-means with smart centers initialization). The feasibility and low cost of the algorithm to process partitions of big datasets allow the wide use of this clustering method oriented to many applications. To define the optimal number of groups of users, given the dataset in analysis we also applied the Elbow Heuristic.  

\section{Experimental Framework}\label{experiments}
In this section, we detail the experimental framework which validates our proposal. We present a case study based on a real-world scenario and have divided the section in the following. First, we describe the datasets employed during the experiments; then, we provide an explanation about the baseline approach used for comparison. Later, the experimental setup followed by the corresponding results are discussed.

\subsection{Data Collection}\label{ data}
To run the experiments and implement our approach we need some datasets: 

\begin{compactitem}
\item a set of tweets to train the word2vec model,
\item a list of users and their tweets/retweets, and
\item a list of users whose profile or preferred topic is well known in order to evaluate the performance of the baseline method and the proposed approach.
\end{compactitem}
The detailed description of the data is provided next.

\subsubsection{Training Corpus to obtain the Vocabulary Model}\label{corpus}
As it was said before, the authors collected datasets with the aim of applying word2vec. The trained model, which was the result of the research done in \cite{dgoRecalde}, was used in the present work because of the advantages the dataset presented: \textit{i)} diverse nature of content because of a pool of 319,889 tweets posted by Ecuadorian users during a month, and \textit{ii)} the authors have knowledge of the context involved, \textit{i.e.} hashtags and their topics, meaning of referenced places and events, and public figures as well as the category their posts fall in. The previous research explored and validated the quality of the training dataset. Indeed, the vocabulary extracted and represented as vectors covers most of the words Ecuadorian tweeters tend to use. Therefore, it suggests that the model can be generalized for similar scenarios as the one presented in this research. Besides, after doing some tests, it was found that the appropriate representation for this kind of input text (short sentences in Spanish) is of 300 dimensions. The trained model corresponds to the output of the approach phase presented in Section \ref{modeling}, Tweets Modeling. Once these tweets are modeled we identified the number of topics involved (Section \ref{suitable}) and the centroids to then initialize EM. Moreover, the vocabulary vectors are later used to define other tweet models.

\subsubsection{Sample of Users and their Timeline}\label{users}
A set of 360 users was sampled from the list of tweeters who created the tweets in section \ref{corpus}. Every tweet in the corpus has meta-data that has information about of it, such as `text' of the tweet, `creation date', `list of hashtags' contained in the tweet, `user' (id number and screen name) who posted the tweet, among others. Given that we have a list of 37,628 users, we had to randomly sample 360 of them due to the Twitter API rate limits. To apply the proposed method, we extracted the last 3,200 tweets from their accounts. Finally, the amount of tweets collected from the users' timelines is of 236,453. 

\subsubsection{Sample of Users for Approach Evaluation}\label{politicians}
We considered a list of 39 political figures who currently participate in the government as public employers or who were candidates for government positions during the 2017 elections. We query their Twitter accounts and extracted a total list of 58,533 tweets. These tweets were added to the set previously obtained. Then, we will apply our approach (Section \ref{em}) considering a dataset of 294,986 tweets in total. It is worth mentioning that those tweets belong to the 399 users. 39 of them are politicians intentionally added to test the accuracy of the proposed approach. In other words, the political figures help us to validate if after getting their \textit{MUMs} and clusters (Sections \ref{mum}, \ref{clustering}), they are going to be found as similar (homogeneous profile models) and put together. In such a case, we can assure that the tweets and users are being correctly modeled. 

\subsection{Baseline Approach}\label{sec:baseline}
To compare the performance of the MUM approach at modeling tweeters, a baseline method is proposed. We elaborate a strategy made of core techniques. What follows is a map of our approach phases and the decisions made to construct the baseline.
\begin{enumerate}
    \item {\bf Tweets Modeling.} The dataset of tweets presented in \ref{users} (training corpus) was modeled by applying \textit{tf-idf} \cite{Baeza-Yates:1999:MIR:553876}. Such a strategy is one of the core information retrieval techniques used to create a vector representation of text.
    \item {\bf Extraction of the suitable Number of Topics.} To build a ground truth about the topics hidden in the tweets dataset and get a subset of classified tweets, we extracted a list of the most frequent hashtags present in the tweets. We inspect the hashtags to identify keywords corresponding to a given category. For example, the hashtags \textit{\#ecu911, \#routesecu911 and \#ecu911withme} lead us to define the topic \textit{Citizens Safety and Emergencies}. As a result, 22 topics were extracted and the corresponding tweets, which contained the studied hashtags, were labeled accordingly. Usually, this manual classification technique allows the categorization of 20\% of the tweets. In our case, from 319,889 tweets we classified 68,186 which corresponds to the 21.3\%. The 22 categories define the number of dimensions the users’ model will have.

\item {\bf Tweets Classification.} In our approach, EM is used to generate a topic-soft-assignment for each tweet (Mixture of Gaussians). For the baseline approach, we will predict the \textit{topic} of the given tweet by applying a traditional machine learning algorithm. We did a series of tests to select an appropriate classification algorithm. First, we chose three machine learning approaches used to realize \textit{multi-class} prediction. These were logistic regression, decision trees and boosting trees. Then, we took 80\% of the previously label tweets to be the training dataset. The rest of the tweets were used to test the models.
\begin{figure}
\centering
\includegraphics[scale=0.6]{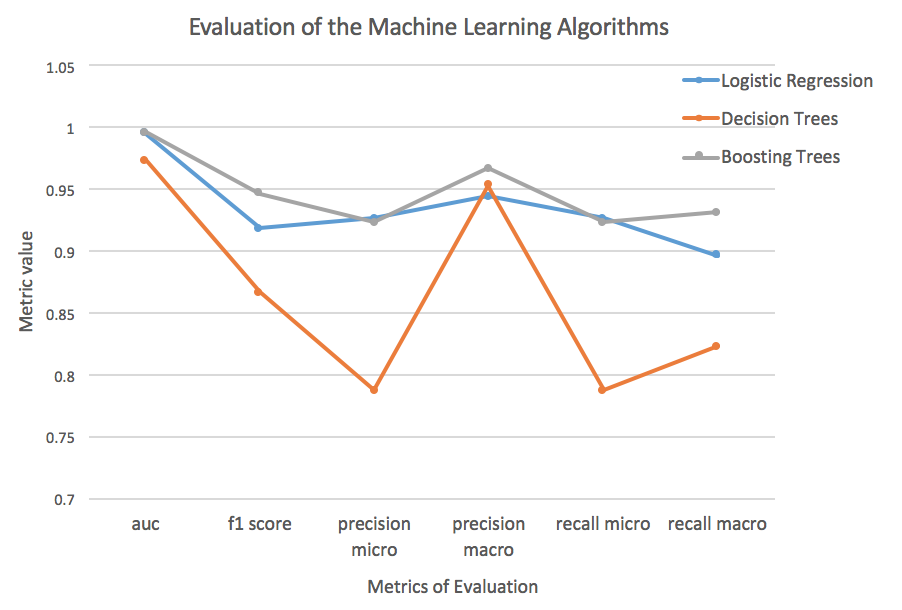}
\caption{Comparison of the performance of the machine learning algorithms (multi-class prediction).}
\label{fig:MachineL}
\end{figure}
As it is shown in Figure \ref{fig:MachineL}, Boosting Trees algorithm \cite{Freund:1996} outperformed the others, so it was used to classify the users' tweets in next phase. The algorithm is based on a technique called gradient boosting, which combines a collection of base learners (i.e. decision tree classifiers) for predictive tasks. It can model non-linear interactions between the features and the target. It is worth clarifying that for precision and recall we calculated the micro and macro values. Micro precision/recall calculates the metrics globally by counting the total true positives, false negatives, and false positives. On the other hand, the macro value calculates the metrics for each label and finds their unweighted mean (label imbalance is not considered). We use the trained boosted trees model to get the class/topic of the new observations (294,986 tweets of the 399 users with their tf-idf representation). As output, we obtain the \textit{class} and the corresponding \textit{class-probabilities}\footnote{\url{https://turi.com/products/create/docs/generated/graphlab.boosted_trees_classifier.BoostedTreesClassifier.classify.html}}.\\

\item {\bf Twitter Users Model.} According to our proposal, the $MUM$ method aggregates the results of the EM algorithm applied over the tweets of a given user. On the other hand, considering the baseline approach, we take the tweets of the target user $T_u$ and their probabilities associated to the class prediction $P_t$ (results of the boosting trees classifier). At last, to define the user's model $M$, we average the probabilities obtained for each of the 22 classes:
\begin{equation*}
M_j = avg(\sum_{i=0}^{|T_u|-1} P_t^{ij})
\end{equation*}
For $j \in [0, 21]$.

At the end of these baseline method's stage, the users will have a set of values (j) that quantify the level of preference of the user for the corresponding topics.\\

\item {\bf Grouping like-minded Users.} We take this phase to evaluate the performance of the baseline approach. In order to compare our method and the baseline, this step was identically applied in both $MUM$ and $M$ (Refer to Section \ref{clustering}). More detailed about the obtained results is given in Section~\ref{sec:validation}.
\end{enumerate}

\subsection{Experimental Setup and Strategy}\label{strategy}
\begin{figure}
\centering
\includegraphics[scale=0.36]{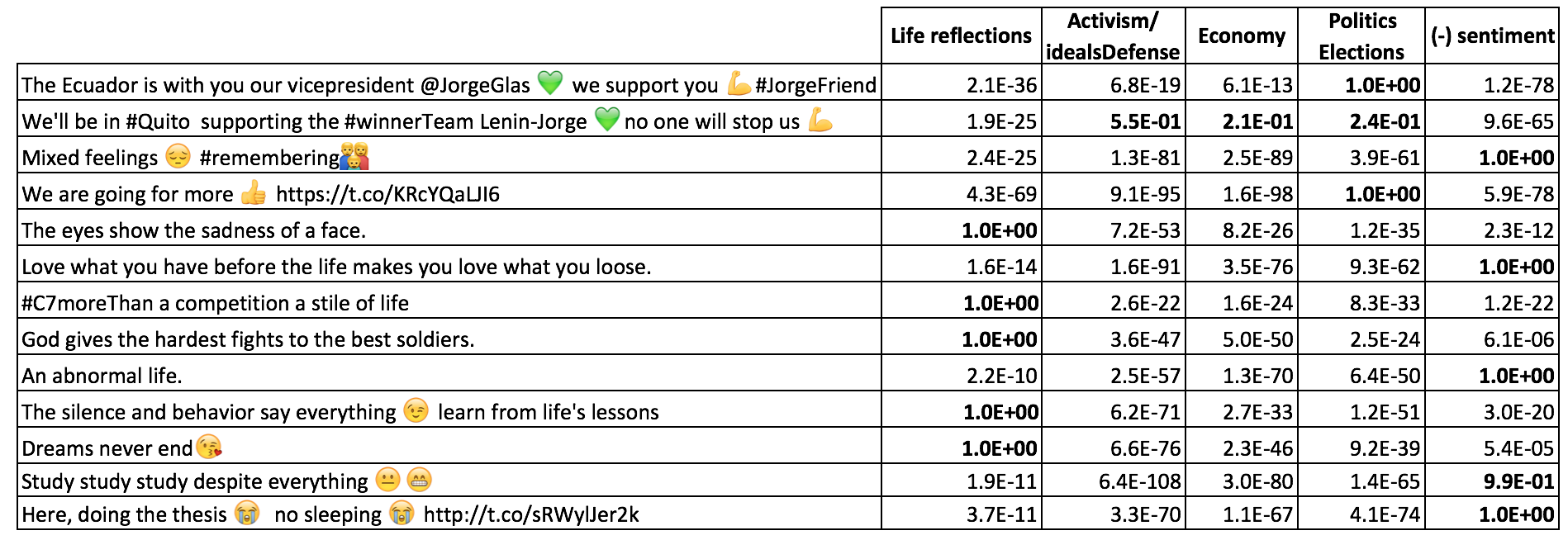}
\caption{Example of Topic assignment with EM algorithm}
\label{fig:example}
\end{figure}

The parameters used to apply word2vec over the \textit{training corpus} are: {\em size=300}, {\em window=5}, {\em sample=0} and {\em min\_count=5}. Other parameters are not modified and take the default values. The output of the word2vec model contains a vocabulary of 39,216 words represented as vectors. Equation \ref{tweet_vector} is applied to have the vectors of the tweets in the training corpus. When the set of $w'_t$ is ready we can move on to the next phase to define the number of clusters in which the tweets are classified.  
\begin{figure}
\centering
\includegraphics[scale=0.6]{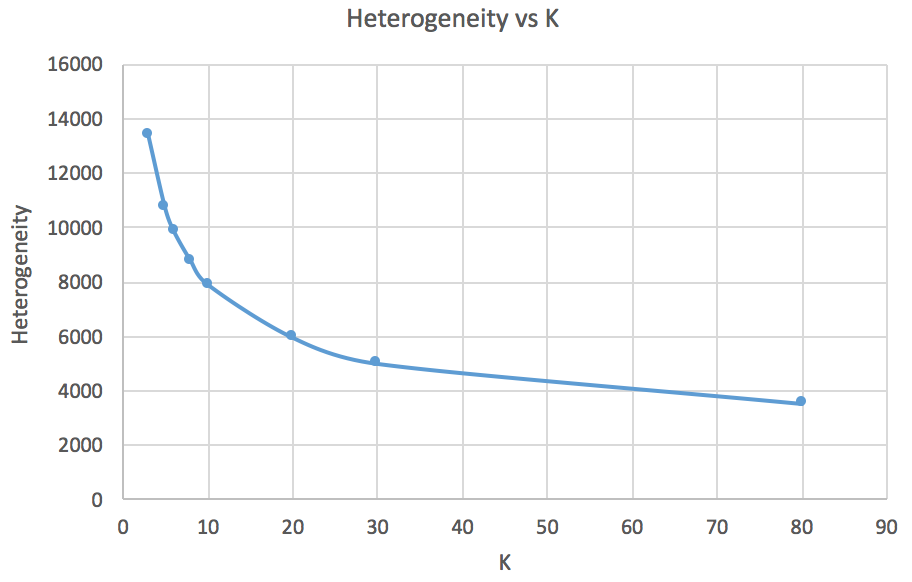}
\caption{Elbow Heuristic: Heterogeneity vs $K$ values.}
\label{fig:Heterogeneity}
\end{figure}
We run some experiments considering $K$ (number of clusters to find) equal to several values. For each given $K$ we apply K-Means++ to cluster the tweets and after that, we will be able to calculate the heterogeneity (Equation \ref{heterogeneity})\footnote{It has to be mentioned that for the given $K$ we run K-Means++ with some initialization seeds: 0, 20000, 40000, 60000, 80000. The considered seed to define the centroids for our work was the one which reported the minimum heterogeneity.}. The results are shown in Figure \ref{fig:Heterogeneity} where we have the heterogeneity vs $K$ plot. The Elbow Heuristic specifies that by analyzing this plot, we can define the optimal number of clusters for the provided data points. The diagram shows that the gain reduces significantly from $K$=3 to $K$=20. Besides, we see a flattening out of the heterogeneity for $K>=30$ (overfitting for larger values of K). So, it might indicate that the $K$ searched is in a range of 20 and 30. To make a decision, we take into account the manual classification of the training tweets in the baseline method, where \textit{22 topics} were found. Whereby, as the Elbow Heuristic also suggests, we consider 22 topics, or $K=22$ to continue working on our approach. The centroids for the 22 clusters are calculated and used to initialize the $means$ for EM. When applying the EM algorithm in order to get a soft topic assignment per tweet, we will be using the dataset of 399 users' tweets (39 of the users are political figures, which results are employed in Section \ref{sec:validation} for validation). The resulting responsibility matrix is used to define the MUM of the users by implementing Equations \ref{userEq1}, \ref{userEq2} and \ref{userEq3}. As an example, Figure \ref{fig:example} shows 5 topics and the degree of responsibility they have over 13 tweets of a given user\footnote{It is worth noting that, as other unsupervised methods, the names of the classes, categories or topics are not defined by the proposed clustering strategy. For the example in Figure \ref{fig:example}, to provide the topic labels, we extracted and analyzed the tweets classified in the corresponding topic with a minimum value of 0.90. Doing so, we were able to annotate the category names.}. The user we took had 698 tweets and once we extracted his/her MUM, the model presented a value of 49.1 for the topic `(-) sentiments' and 11.4 in `life reflections' (highest category weights).
The model of tweeters is finally obtained and may be used with many purposes. Actually, to align the results with the goals of our research we cluster the users to define groups of tweeters with \textit{similar profiles} or tastes about content topics (last phase of our approach, Section \ref{clustering}). By making use of the notion about heterogeneity and Elbow Heuristic we find that the users in our dataset form 5 clusters. To evaluate the behavior of our approach facing the chosen baseline, we introduced a set of politicians. The assumption behind this is that if their profile is well represented, they are going to be grouped in the same cluster. This validation is presented in next Section.

\subsection{Validation of Results}\label{sec:validation}
The users we take to do this validation are well-known political figures who have a position in the government or were candidates in different democratic elections. The clustering algorithm we applied with the aim of validating the $MUM$ approach as well as the results of the baseline method was K-Means++. The details about the results for both approaches are presented in Table \ref{clusterUsers}.
\begin{table}
\caption{Summary of Users Clusters: Baseline and MUM methods.}
\label{clusterUsers}
\begin{center}
\scriptsize
\begin{tabular}{|c|m{1.3cm}|m{1.3cm}|m{1.8cm}|m{1.8cm}|}
\cline{2-3}
\hline
Cluster ID & Total Size (Baseline) & Total Size ($MUM$) & Politicians Classification (Baseline) & Politicians Classification ($MUM$)\\
\hline \hline
0 & \multicolumn{1}{r|}{50}  & \multicolumn{1}{r|}{100} & \multicolumn{1}{r|}{17}  & \multicolumn{1}{r|}{36} \\ \cline{1-5}
1 & \multicolumn{1}{r|}{165} & \multicolumn{1}{r|}{6} & \multicolumn{1}{r|}{0}  & \multicolumn{1}{r|}{0} \\ 
\cline{1-5}
2 & \multicolumn{1}{r|}{126} & \multicolumn{1}{r|}{45} & \multicolumn{1}{r|}{0}  & \multicolumn{1}{r|}{1} \\ 
\cline{1-5}
3 & \multicolumn{1}{r|}{16} & \multicolumn{1}{r|}{122} & \multicolumn{1}{r|}{2}  & \multicolumn{1}{r|}{1} \\ 
\cline{1-5}
4 & \multicolumn{1}{r|}{43} & \multicolumn{1}{r|}{127} & \multicolumn{1}{r|}{20}  & \multicolumn{1}{r|}{1} \\
\hline 
\end{tabular}
\end{center}
\end{table}
The Table also shows how the politicians were classified. In the case of the baseline implementation, we can see that there are two prominent groups of politicians. One group covers 44\% of them, while the other groups the 51\%. By analyzing the centroids of the two clusters, we identified that \textit{cluster 4}, differently from \textit{cluster 0}, groups users who tend to talk more about economy. Compared to our approach, it is shown that MUM performance at clustering politicians has 92\% of precision. From the 39 politicians, only 3 were left out of the political-related cluster. The `screen\_name' of these users are $lcparodi$, $ramiroaguilart$ and $mmcuesta$. By verifying their MUM (the 22 dimensions of the model) and their tweets, it is seen that their profiles are different from the rest of politicians who mostly talk about elections, economy and social issues. Instead, lcparodi tweeted about capital market and investment, ramiroaguilart posted about his interviews in radio media and talks directly to people loading his account of mentions ($@$); besides, our model separated mmcuesta because she talks about recipes/food and cooking, and she promotes few enterprises\footnote{We have to mention that lcparodi  and mmcuesta are the users who belong to cluster 3 in the Politicians Classification - Baseline Method.}. 

\begin{figure}
\centering
\includegraphics[scale=0.36]{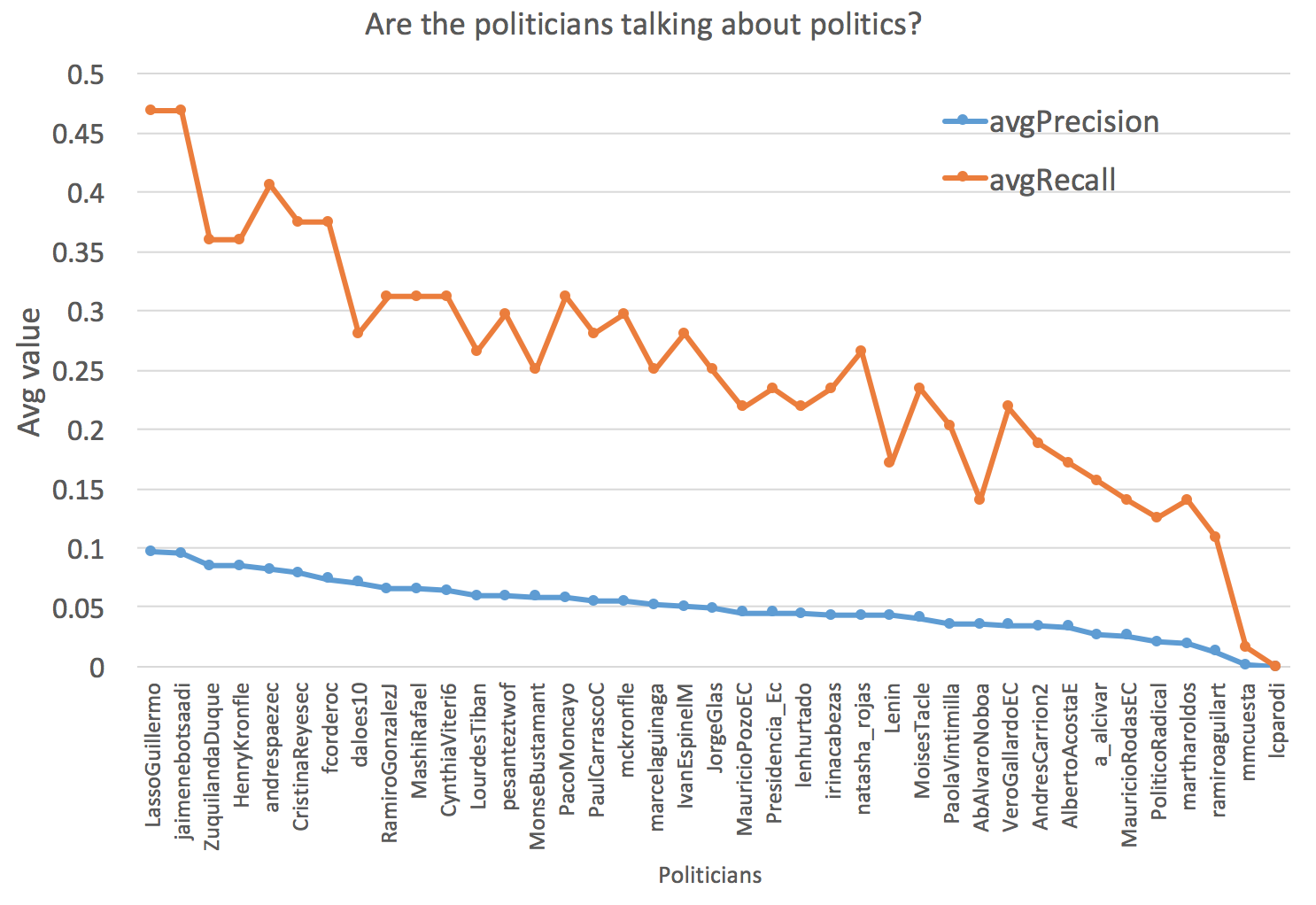}
\caption{Relevance of the context of ``Politics'' in the politicians' Twitter accounts}
\label{fig:politicians}
\end{figure}

In order to make a deeper comparison of the politicians who were clustered together and the rest three, we did text mining over their Twitter accounts. As we already collected their time-lines, we consider every politician's tweets as a document, \textit{i.e.} there is a collection of 39 documents to be analyzed. We apply tf-idf over this corpus and found the most relevant words for the corresponding politicians' profiles. From among the most frequent words in the whole corpus, a list of meaningful words in the context of ``politics'' was extracted. The mentioned list contains the words: Ecuador, government, country, Ecuadorians, president, `the people' (pueblo), job, work, city, production, laws, taxes, congress, health, justice, and citizens. In this experiment we try to find if the previous list was present among the relevant words extracted for the politicians. We worked with the 30, 50, 100 and 200 most relevant words taken from their profiles. The results for the average precision and recall are explained in Figure \ref{fig:politicians}. 
As it is showed, the users $ramiroaguilart$, $mmcuesta$ and $lcparodi$ have the minimum values for both precision and recall; then, it is proved that they did not discuss about political issues as the rest of the politicians.

\section{Conclusions and Future Work}\label{conclusions}
People may show preference for several topics to a greater or lesser extent. In this research, we have proposed a method that creates a vector representation of tweets by applying word2vec. Then, by using a Mixture of Gaussians through the Expectation Maximization algorithm, it calculates the degree of responsibility that a set of topics have over a tweet. Finally, we aggregate the results of the tweets which correspond to a given user to define his/her multi-topic preference model. We have validated our proposal by comparing it with the results of a baseline approach. This evaluation showed that our method was able to cluster 92\% of politicians in the same group, facing the results of the baseline method which divided the politicians in two clusters. In summary, we can conclude that our method is effective when modeling the topic interests of Twitter users.

There are other issues to be discussed. In fact, the authors think that the most important step is the definition of the training dataset. Then, for future work, we consider updating the vocabulary obtained with word2vec algorithms due to new topics/hashtags appear over time. We think that our method can be used in recommender systems to find new content and subscription lists that match the users’ profiles. We propose for further research to label the groups of users and to apply a validation not only by identifying the group of politicians but also other clusters of users modeled with MUM. Also, we plan to evaluate our approach with other probabilistic topic models like LDA and test its performance at topic assignment for short text.

\end{document}